\documentstyle[12pt,epsfig]{article}

\newcommand{\be}{\begin{equation}}
\newcommand{\ee}{\end{equation}}
\def\bea{\begin{eqnarray}}
\def\eea{\end{eqnarray}}
\def\nn{\nonumber \\}

\def\o{\omega}
\def\e{\epsilon}

\input psfig

\begin{document}

\date{}
\title{
{\large\rm DESY 97-083}\hfill{\large\tt ISSN 0418-9833}\\
{\large\rm May 1997}\hfill\vspace*{2.5cm}\\
Classical Statistical Mechanics and Landau Damping}
\author{W. Buchm\"uller and A. Jakov\'ac\\
{\normalsize\it Deutsches Elektronen-Synchrotron DESY, 22603 Hamburg, Germany}
\\[.2cm]
\vspace*{2cm}\\                     
}                                                                          


\maketitle  
\begin{abstract}
\noindent
We study the retarded response function in scalar $\phi^4$-theory
at finite temperature. We find that in the high-temperature limit the
imaginary part of the self-energy is given by the classical theory to
leading order in the coupling. In particular the plasmon damping rate
is a purely classical effect to leading order, as shown by Aarts and Smit.
The dominant contribution to Landau damping is given by the propagation of
classical fields in a heat bath of non-interacting fields. 
\end{abstract} 
\thispagestyle{empty}
\newpage                                             

In the high-temperature phase of the standard model of electroweak 
interactions baryon-number and lepton-number violating processes are in 
thermal equilibrium \cite{kuz}. This is of crucial importance for the 
presently observed cosmological baryon asymmetry and the `sphaleron rate' 
of baryon-number violation is a corner stone of the theory of baryogenesis.

In order to obtain the sphaleron rate one has to compute a real-time
correlation function at finite temperature in a non-abelian gauge theory -
a problem which, up to now, could not be solved. Several years ago Grigoriev
and Rubakov suggested that the dominant contribution to the sphaleron rate
could be obtained by computing the classical time evolution of the gauge
fields and averaging over the initial conditions with the Boltzmann weight
factor \cite{gri}. This classical real-time method has indeed led to an 
accurate determination of the `classical' sphaleron rate which was shown to 
be insensitive to the ultraviolet cutoff \cite{amb}. The same procedure has
been used to study other real-time properties of the SU(2)-Higgs model at
finite temperature \cite{moo,tan}. 

The classical real-time method has been questioned for several reasons. One 
worry concerns the well know ultraviolet divergencies of classical statistical
field theory \cite{smi,boe}. Another open problem is the role of Landau
damping \cite{mcl,arn}, i.e., the loss of energy of quasi-particle
excitations in the plasma to the heat bath.

To elucidate the role of classical field theory for finite-temperature
quantum field theory, Aarts and Smit have recently studied the plasmon
damping rate in scalar $\phi^4$-theory \cite{aar}. As they have shown, 
the damping rate is determined by the classical theory to leading order in the
coupling. In this paper we extend this analysis by studying the retarded 
response function in scalar $\phi^4$-theory. It turns out that the imaginary
part of the self-energy is entirely given by the classical 
theory in the high-temperature limit to leading order in the coupling. 
We also calculate the dominant subleading contribution to the plasmon 
damping rate which is a genuine quantum effect.\\

\noindent
{\bf\large Retarded Response Function in the Quantum Theory}\\

We consider scalar $\phi^4$-theory at temperature $T=1/\beta$. Perturbation
theory is most easily carried out in the imaginary time formalism which is
based on the action
\bea
S &=&\int_{\beta}dx \left({1\over 2} \partial_{\mu}\phi\partial_{\mu}\phi 
+ {1\over 2}m^2\phi^2 + {1\over 4!}\lambda\phi^4 + \delta\right)\, ,\nonumber\\
&&\qquad \int_{\beta}dx = \int_{0}^{\beta}d\tau\int d^3x\, .
\eea
Here $m$ is the plasmon mass, 
\be\label{mplas}
m^2 = {1\over 24}\lambda T^2\, ,
\ee
and $\delta$ stands for counter terms associated with zero-temperature
ultraviolet divergencies as well as the resummation at finite temperature
\cite{dol}. The two-point function
\be
D(\tau_1-\tau_2,\vec{x}_1-\vec{x}_2) = \langle \phi(\tau_1,\vec{x}_1)
\phi(\tau_2,\vec{x}_2)\rangle
\ee
satisfies the periodic boundary condition $D(0,\vec{x})=D(\beta,\vec{x})$.
Hence, its Fourier transform is only defined for the discrete energies
given by the Matsubara frequencies $\omega_n = 2\pi n/\beta$, with
$n=0,\pm 1, \pm2, \ldots$,
\be
D(\tau,\vec{x}) = {1\over \beta}\sum_n e^{i \o_n\tau}
\int d^3x e^{i\vec{p}\vec{x}} D(i\o_n,\vec{p})\, .
\ee
Analytic continuation yields the function $D(z,\vec{p})$ for complex 
values of $z$ \cite{wel}. This function is related to the self-energy by 
a Dyson-Schwinger equation,
\be
\Pi(z,\vec{p}) = D^{-1}(z,\vec{p}) - D^{-1}_0(z,\vec{p})\, .
\ee
Here $D_0(z,\vec{p})=1/(z^2-\o_p^2)$, with $\o_p=\sqrt{\vec{p}^2+m^2}$, 
is the free two-point function. Along the real axis real and imaginary part
of the self-energy are defined by ($\o=\o^*$),
\be
\Pi(\o\pm i\epsilon,\vec{p}) = \Pi_R(\o,\vec{p}) \pm
        i\ \Gamma(\o,\vec{p})\, .
\ee

Knowing $D(z,\vec{p})$ in the complex plane, the various two-point
functions can be evaluated by choosing the appropriate path. We are
particularly interested in the retarded two-point function which is
given by
\be\label{dr}
D_R^{qu}(t,\vec{p})=\int_{-\infty}^{+\infty}{d\o\over 2\pi} 
D(\o + i\epsilon, \vec{p}) e^{-i \o t}\, .
\ee
For small couplings $\lambda$ the self-energy $\Pi(z,\vec{p})$ can be 
evaluated in perturbation theory. If $D(z,\vec{p})$ has a pole at 
\be
\bar{z} = \bar{\o}_p - i \bar{\gamma}_p\, ,
\ee
with
\be
\bar{\o}_p \simeq \sqrt{\o_p^2 - \Pi_R(\o_p,\vec{p})}\, , \quad
\bar{\gamma}_p \simeq {1\over 2\o_p} \Gamma(\o_p,\vec{p})\, ,
\ee
the integral in Eq.~(\ref{dr}) is easily evaluated, and one obtains the
ordinary retarded Green's function,
\be
D_R^{qu}(t,\vec{p}) \simeq -\Theta(t) {1\over \bar{\o}_p} \sin(\bar{\o}_p t)
                 e^{-\bar{\gamma}_pt}\, .
\ee

The retarded two-point function can also be expressed as ensemble average of
a product of field operators,
\be \label{drqu}
i D_R^{qu}(t-t',\vec{x}-\vec{x}') = \mbox{Tr}\left(e^{-\beta \hat{H}}
[\hat{\phi}(t,\vec{x}),\hat{\phi}(t',\vec{x}')]\right) \Theta(t-t')\, ,
\ee 
where $\hat{H}$ is the hamilton operator of the theory. Hence, $D_R$
describes the response of the system to a perturbation
caused by an external current. Adding to the hamiltonian a source term,
\be
\hat{H}_s(t) = \int d^3x  j(t,\vec{x}) \hat{\phi}(t,\vec{x})\, ,
\ee
leads to a non-vanishing ensemble average of the field operator. In linear 
response theory one obtains
\bea\label{linrep}
\langle\hat{\phi}(t,\vec{x})\rangle_{qu} &=&
\mbox{Tr}\left(e^{-\hat{H}}\hat{\phi}(t,\vec{x})\right) \nn
&=& \int dt'\int d^3x' j(t',\vec{x'})D_R(t-t',\vec{x}-\vec{x}')\, .
\eea

The finite-temperature two-point functions have been studied in detail in
the literature. The self energy $\Pi(z,\vec{p})$
has been calculated in perturbation theory to two-loop order \cite{par,hei}.
We will be particularly interested in the damping rate, i.e., the imaginary
part. It consists of two parts \cite{wel}: the first one, which is present
also at zero temperature, is due to the off-shell decay of a scalar particle 
into three on-shell scalar particles; the second part results from
Landau damping, i.e., the scattering of the off-shell scalar particle
from the heat bath. The imaginary part of the self-energy reads at two-loop
order ($\o > 0$) \cite{hei},
\bea
\Gamma(\o,\vec{p})&=&{\pi\over 6}\lambda^2 \left(e^{\beta\o}-1\right)
\int d\Phi(\vec{p}) \left[\delta(\o-\o_1-\o_2-\o_3) f_1 f_2 f_3\right. \nn
 && \hspace*{3cm} +\left. 3 \delta(\o+\o_1-\o_2-\o_3) (1+f_1)f_2 f_3 
     \right]\, ,\label{gamq}\\
\Gamma(-\o,\vec{p}) &=& - \Gamma(\o,\vec{p})\, ,
\eea
where $d\Phi(\vec{p})$ is the integration measure,
\be
d\Phi(\vec{p}) = {d^3q_1\over (2\pi)^3 2\o_1}
 {d^3q_2\over (2\pi)^3 2 \o_2} {d^3q_3\over (2\pi)^3 2 \o_3}
 (2\pi)^3 \delta(\vec{p}-\vec{q_1}-\vec{q_2}-\vec{q_3}) \, ,
\ee
and $f$ is the Bose-Einstein distribution function,
\be
f(\o) = {1\over e^{\beta \o } - 1}\, .
\ee
The two contributions to $\Gamma(\o,\vec{0})$ have thresholds at $\o=3 m$
and $\o= m$, respectively. Only the first term is present at zero temperature.
At temperatures $T \gg m$ both terms are dominated by interactions with
the heat bath, i.e., by Landau damping.\\

\noindent
{\bf\large Retarded Response Function in the Classical Theory}\\ 

The retarded response function can also be studied in the classical theory
at finite temperature. The classical hamiltonian is the sum of the
free hamiltonian, a self-interaction term and a source term,
\be
H = \int d^3x \left( {1\over 2} \pi^2 + {1\over 2} (\vec{\partial}\phi)^2
    + {1\over 2} m^2_{cl} \phi^2 + {1\over 4!}\lambda \phi^4 
    + j \phi \right)\, ,
\ee
where $j(t,\vec{x})$ is an external souce. Note, that $m_{cl}=1/l_{cl}$ 
is an inverse length. It is different from the plasmon mass $m$ given in
Eq.~(\ref{mplas}). The ensemble average of the classical field is then given by
\be\label{fi1}
\langle \phi(t,\vec{x})\rangle_{cl} = {1\over Z} 
\int {\cal D}\pi {\cal D}\phi e^{-H(\pi,\phi)}\phi(t,\vec{x})\, ,
\ee
with
\be\label{fi2}
Z = \int  {\cal D}\pi {\cal D}\phi e^{-H(\pi,\phi)}\, .
\ee
Here $\phi(t,\vec{x})$ is the solution of the equations of motion
\bea
\dot{\phi}(t,\vec{x}) &=& \pi(t,\vec{x})\, ,\nn
\dot{\pi}(t,\vec{x}) &=& \Delta\phi(t,\vec{x}) - m^2_{cl} \phi(t,\vec{x})
             - {\lambda\over 3!} \phi^3(t,\vec{x}) - j(t,\vec{x})\,  ,
\eea
with the initial conditions
\be
\phi(t_0,\vec{x}) = \phi(\vec{x})\, , \quad
\pi(t_0,\vec{x}) = \pi(\vec{x})\, .
\ee
In Eqs.~(\ref{fi1}) and (\ref{fi2}) the domain of integration is the space
of initial conditions.

The solution of the equations of motion satisfies the integral equation
($t,t'> t_0, x\equiv (t,\vec{x})$),
\be\label{phj}
\phi(x;j) = \phi_0(x) + \int d^4x' D_R(x-x')\left(
  {\lambda\over 3!} \phi^3(x';j) + j(x')\right)\, .
\ee
Here $\phi_0$ is the solution of the free equations of motion which
satisfies the boundary condition, and $D_R$ is the retarded Green's function,
\be\label{dcl}
D_R(t,\vec{x}) = -\int {d^3q\over (2\pi)^3} e^{i\vec{q}\vec{x}}
  \Theta(t){1\over \o_q}\sin(\o_q t)\, .
\ee
 
The classical solution can be expanded in powers of the external current.
The term linear in $j(x)$ is given by (cf.~(\ref{linrep}))
\be\label{phi}
\phi(x) = \int d^4x' j(x') H_R(x-x')\, ,
\ee
where 
\be\label{hrd}
H_R(x-x') = \left.{\delta \phi(x;j)\over \delta j(x')}\right|_{j=0}
\ee
depends on the initial conditions $\phi(\vec{x})$ and $\pi(\vec{x})$.
From Eqs.~(\ref{phj}) and (\ref{hrd}) one reads off the integral equation 
which determines $H_R$,
\be\label{hr}
H_R(x-x') = D_R(x-x') + {1\over 2}\lambda \int d^4y D_R(x-y) 
                        \phi^2(y;0)H_R(y-x')\, .
\ee
In order to obtain the finite-temperature retarded response function, i.e.,
the classical analogue of $D_R^{qu}$ (cf.~(\ref{drqu})), one has to compute
the ensemble average with respect to the initial conditions. This yields
\be
\langle \phi(x)\rangle_{cl} = 
\int d^4x' j(x') D_R^{cl}(x-x')\, , 
\ee
where
\be
D_R^{cl}(x-x') = \langle H_R(x-x') \rangle_{cl} =
 {1\over Z} \int {\cal D}\phi {\cal D}\pi  e^{-\beta H} H_R(x-x')\, .
\ee

 The classical retarded response function $D_R^{cl}$ can be evaluated
by first expanding the solution (\ref{hr}) for $H_R$ in powers of
$\lambda$ (cf.~Fig.~1) and then performing the thermal average for each term.
\begin{figure}[htbp]
  \begin{center} 
  \leavevmode 
     \epsfig{file=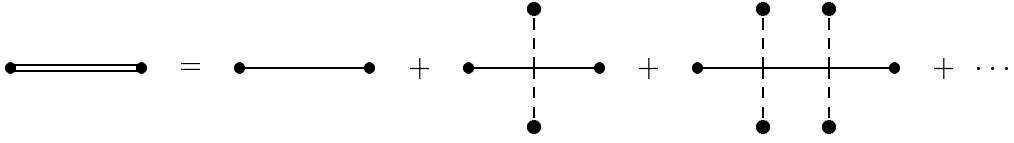} 
  \end{center}
  \caption{\em The response function $H_R$ expanded in powers of the
     classical field. Full lines denote retarded Green's functions
     $G_R$}
\end{figure}
This requires the evaluation of the thermal n-point functions
\be
\langle \phi^2(x_1) \ldots \phi^2(x_{2n})\rangle_{cl}\, .
\ee
Consider first the thermal average with the free Hamiltonian $H_0$. The
corresponding two-point function reads \cite{gpa,aar}
\be\label{scl}
S(t-t',\vec{x}-\vec{x}') = \langle \phi(x) \phi(x') \rangle_{cl}^0 \nn
= T \int {d^3q\over (2\pi)^3} e^{i\vec{k}(\vec{x}-\vec{x}')}
    {1\over \o_k^2} \cos(\o_k(t-t'))\, .
\ee
For higher n-point functions one has
\be
\langle \phi^2(x_1) \ldots \phi^2(x_{2n})\rangle_{cl}^0 = 2^n
S^2(x_1-x_2) \ldots S^2(x_{2n-1}-x_{2n}) + \mbox{permutations}\, .
\ee

From Fig.~1 it is clear that the response function obtained after thermal 
averaging with $H_0$ satisfies a Dyson-Schwinger equation (cf.~Fig.~2),
\be
\bar{D}_R^{cl}(x-x') = D_R(x-x') + \int d^4y d^4y' 
        D_R(x-y)\bar{\Pi}(y-y')_{cl} \bar{D}_R^{cl}(y'-x')\, . 
\ee
\begin{figure}[htbp]
  \begin{center} 
  \leavevmode 
     \epsfig{file=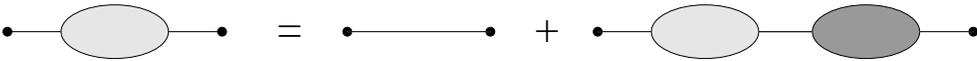,width=13cm} 
  \end{center}
  \caption{\em Dyson-Schwinger equation for the retarded response
     function $\bar D_R^{cl}$}
\end{figure}
The terms in the perturbative expansion for the self-energy $\Pi(x)$ are
shown in Fig.~3 up to ${\cal O}(\lambda^2)$.
\begin{figure}[htbp]
  \begin{center} 
  \leavevmode 
     \epsfig{file=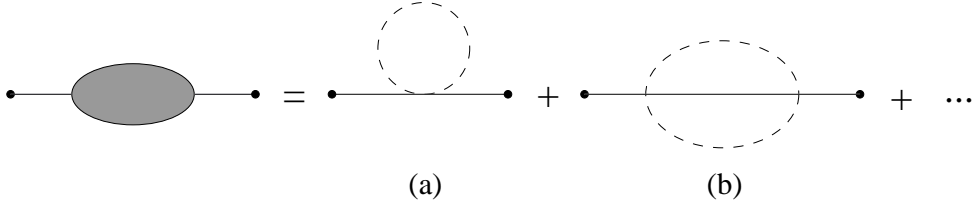, width=13cm} 
  \end{center}
  \caption{\em Perturbative expansion for the self-energy
     $\Pi$. Dashed lines denote thermal two-point functions $S$.}
\end{figure}
The contribution 
${\cal O}(\lambda)$ (Fig.~3a) is linearly divergent and can be removed 
by a mass renormalization yielding the counter term \cite{aar}
\be
\delta m^2_1 = {1\over 2}\lambda S(0,\vec{0}) = {1\over 2}\lambda T 
               \int {d^3q\over (2\pi)^3} {1\over \omega_q^2}\, .
\ee
The contribution ${\cal O}(\lambda^2)$ (Fig.~3b) is logarithmically 
divergent and can be rendered finite by a further mass renormalization
\be
\delta m^2_2 = -{1\over 6}\lambda^2\beta\int d^3x S^3(0,\vec{x})
  = -{4\over 3}\lambda^2 T^2 \int d\Phi(\vec{0}){1\over \o_1\o_2\o_3}\, .
\ee
Contributions of higher order in $\lambda$ are all finite.

We are interested in the damping rate, i.e., the imaginary part
of the self-energy which is finite. Using Eqs.~(\ref{dcl}) and (\ref{scl}) 
one easily obtains for the self-energy to order $\lambda^2$,
\bea\label{se}
\bar{\Pi}(\o,\vec{p})_{cl} &=& \int dt d^3x\ e^{i\o t} e^{-i\vec{p}\vec{x}}\
                      \bar{\Pi}(t,\vec{x})_{cl} \nn
&=& {1\over 2} \lambda^2 T^2 \sum_{\eta_1,\eta_2,\eta_3} \eta_1
\int d\Phi(\vec{p})\ {1\over \o_2\o_3}
   {1\over \o + \eta_1\o_1 + \eta_2\o_2 + \eta_3\o_3 + i\e}\, ,
\eea 
where $\eta_i = \pm 1$, $i=1,\ldots,3$.

From Eq.~(\ref{se}) one obtains for the imaginary part of the self-energy
($\o > 0$),
\be\label{gamc}
\bar{\Gamma}(\o,\vec{p})_{cl}={\pi\over 2}\lambda^2 T^2 \int d\Phi(\vec{p}) 
  {1\over \o_1\o_2\o_3} 
  \left[ \o_1\delta(\o - \o_1 - \o_2 - \o_3) +
         \o\delta(\o + \o_1 - \o_2 -\o_3) \right]\, .
\ee

We can now compare the classical damping rate $\bar{\Gamma}_{cl}$ with the
damping rate $\Gamma$ of the full quantum theory as given by
Eq.~(\ref{gamq}). At very high temperatures, i.e., $T\gg m$, the Bose-Einstein
distribution function becomes
\be
f(\o) \simeq {T\over \o} \gg 1\, .
\ee 
Inserting this approximation into Eq.~(\ref{gamq}) yields precisely 
Eq.~(\ref{gamc}). Hence, in the high-temperature limit, where number
densities are very high, the damping rates are given by the classical
theory to leading order. In particular the on-shell plasmon damping
rate reads
\be
\gamma = {1\over 2} l_{qu} \Gamma(m,\vec{0}) \nn
       = {1\over 1536\pi} l_{qu} \lambda^2 T^2\, ,   
\ee
where
\be
l_{qu} = \hbar \left({\hbar\lambda\over 24} T^2\right)^{-1/2}
\ee
is the Compton wave length associated with the plasmon mass (\ref{mplas}).
Replacing $l_{qu}$ by $l_{cl}$ yields the classical plasmon rate. This is
the result of Aarts and Smit \cite{aar}. As our analysis shows, this result
extends to the full off-shell imaginary part of the self-energy in the
high-temperature limit.\\

\noindent
{\bf\large Use of the Classical Theory}\\

So far we have performed the thermal average in the classical theory with 
respect to the free hamiltonian $H_0$, and we have calculated the self-energy
$\bar{\Pi}_{cl}$ to order $\lambda^2$. Higher orders in the expansion
of $H_R$ (cf.~Fig.~1) and, furthermore, the thermal average with the
full hamiltonian $H$ rather than $H_0$ yield corrections of higher
order in $\lambda$.  

However, if one is interested in the classical theory as a tool to 
evaluate the leading contribution of the quantum theory, the full result
of the classical theory is irrelevant. Consider again the two-loop
expression (\ref{gamq}) for $\Gamma(\o,\vec{p})$. In the high-temperature
limit terms proportional to the product of two Bose-Einstein distribution
functions $f(\o_i) f(\o_j)$ dominate. This contribution is given by the
classical theory. Contributions to $\Gamma(\o,\vec{p})$ proportional to a
single distribution function $f(\o_i)$ are subdominant. Such terms do not
occur in the classical theory. They lead to the following correction for the 
on-shell plasmon damping rate,
\be
\delta\Gamma(m,\vec{0}) = \pi\lambda T \int d\Phi(\vec{0}) {1\over \o_3}
                     \delta(m + \o_1 - \o_2 - \o_3)\, ,
\ee
which yields
\be
\delta \gamma = {1\over 512\pi^3} \hbar\lambda^2 \ln{(\hbar\lambda)}T\, .  
\ee
On dimensional grounds such a correction cannot appear in the classical
theory.

In summary, we have shown that in the high-temperature limit the imaginary 
part of the self-energy is determined by the classical theory to leading
order in the coupling, whereas the dominant subleading contribution is
a genuine quantum effect. In order to obtain the leading order contribution
to damping rates it is sufficient to perform the thermal average with
respect to the initial conditions with the free hamiltonian. Such a 
simplification may be useful in numerical simulations for damping rates.   

We would like to thank D.~B\"odeker, A.~Patk\'os and J.~Polonyi for valuable
discussions and comments.

\end{document}